\documentclass[a4paper]{report}

\pdfoutput=1

\usepackage[utf8]{inputenc}
\usepackage[T1]{fontenc}
\usepackage{RJournal}
\usepackage{amsmath,amssymb,array}
\usepackage{booktabs}

\usepackage{changepage}

\begin{document}

%% do not edit, for illustration only
\sectionhead{Contributed research article}
\volume{XX}
\volnumber{YY}
\year{20ZZ}
\month{AAAA}

%% replace RJtemplate with your article
\begin{article}
  % !TeX root = RJwrapper.tex
\title{BCEA: An R Package for Cost-Effectiveness Analysis}
\author{by Nathan Green, Anna Heath, Gianluca Baio}

\maketitle

\abstract{
We describe in detail how to perform health economic cost-effectiveness analyses (CEA) using the R package \textbf{BCEA} (Bayesian Cost-Effectiveness Analysis).
CEA consist of analytic approaches for combining costs and health consequences of intervention(s). These help to understand how much an intervention may cost (per unit of health gained) compared to an alternative intervention, such as a control or status quo. For resource allocation, a decision maker may wish to know if an intervention is cost saving, and if not then how much more would it cost to implement it compared to a less effective intervention.

Current guidance for cost-effectiveness analyses advocates the quantification of uncertainties which can be represented by random samples obtained from a probability sensitivity analysis or, more efficiently, a Bayesian model.
\textbf{BCEA} can be used to post-process the sampled costs and health impacts to perform advanced analyses producing standardised and highly customisable outputs. We present the features of the package, including its many functions and their practical application. \textbf{BCEA} is valuable for statisticians and practitioners working in the field of health economic modelling wanting to simplify and standardise their workflow, for example in the preparation of dossiers in support of marketing authorisation, or academic and scientific publications.
}

\section{Introduction and motivation}
Health economic cost-effectiveness analyses (CEA) consist of analytic approaches for comparing costs and health consequences of alternative interventions.
CEA is focused on assessing how to treat individuals with a given disease or health state.
For resource allocation with CEA, a decision maker may wish to know if an intervention is cost saving, and if not then how much more would it cost to implement it compared to a less effective intervention.
For example, a CEA can help a health commissioner decide which cancer drug regimen to invest in by identifying the option that provides the lowest cost per quality-adjusted life-year (QALY).

It is mandated by many health technology assessment (HTA) agencies internationally, including the UK's National Institute for Health and Care Excellence (NICE), that CEA include the quantification of uncertainties in model parameters. These can be represented by random samples obtained from probability sensitivity analysis (PSA) or, more efficiently, a Bayesian model. Uncertainties in the true value of model inputs are propagated through the CEA model to produce a random sample of costs and health consequences. 

\textbf{BCEA} is a tool for interpreting and presenting the random sample of results from a CEA in a simple, powerful and standardised way, with useful, technically advanced measures and graphical summaries.
\textbf{BCEA} was primarily written to use posterior distribution samples from a Bayesian model (e.g.~run in WinBUGS or Stan) but can take any PSA random samples as inputs. \textbf{BCEA} also aims to be used in a health economic modelling workflow, meaning that it can be plugged-in as one of the steps in a CEA analysis.
\textbf{BCEA} does not provide modelling functionality, like some other CEA packages mentioned below, but the package philosophy (borrowed from UNIX) is to do one thing well by focusing on the analysis following a model run.
That said, it is meant to be extensible and flexible. Currently, \textbf{BCEA} has base R, \textbf{ggplot2} and \textbf{plotly} versions of the plotting functions.
The code is written so that computation of new statistics and new plotting functionality can be easily added. In \textbf{BCEA} the workflow centres around the \texttt{bcea()} function rather than separate functions for each type of statistic, with the aim to reduce the learning curve and easily expose the package functionality.
Finally, \textbf{BCEA} has an expansive suite of functions from basic cost-effectiveness analyses, e.g.~increment benefit (IB) and ICER calculation and plotting, to more sophisticated methods, e.g.~Expected Value of Perfect Partial Information.

The breadth of models used for CEA is wide and growing in complexity and applications \citep{Krijkamp2018, Krijkamp2019}, but their implementation and, in particular, post-processing of their output can (and should be) standardised \citep{Alarid-Escudero2019a}. This has the benefit of greater reliability, facilitating assessment and reuse. Decoupling the modelling from the post-processing allows for flexibility in the CEA model but, as long as its output is in a standard format, then \textbf{BCEA} can be used. Thus, methodologies in CEA modelling can advance independent of the post-processing and presentation. 

For further, in-depth details about \textbf{BCEA} we encourage the package user to consult  \citep{Baio:2013} and \citep{Baio2017}.

\section{Related packages}
There are some packages available on CRAN which provide related functionality to \textbf{BCEA}, albeit with some crucial differences.

The package \textbf{hesim} \citep{Incerti2021} provides a modular and computationally efficient R package for parameterizing, simulating, and analyzing health economic simulation models. The package supports cohort discrete time state transition models, N-state partitioned survival models, and individual-level continuous time state transition models, encompassing both Markov (time-homogeneous and time-inhomogeneous) and semi-Markov processes. Decision uncertainty from a cost-effectiveness analysis is quantified with standard graphical and tabular summaries of a probabilistic sensitivity analysis.

\textbf{heemod} \citep{Filipovic-Pierucci2017} is an implementation of the modelling and reporting features described in common reference textbooks. It allows deterministic and probabilistic sensitivity analysis, heterogeneity analysis, time dependency on state-time and model-time (semi-Markov and non-homogeneous Markov models), etc.
\textbf{heemod} provides a general purpose framework for developing Markov cohort models. 

\textbf{heRomod2} (which is an update of \textbf{heRoMod}), created by Policy Analysis Inc., is not yet available on CRAN but is a package of note. This was originally forked from \textbf{heemod} and is the computational engine of the web app \textbf{h\={e}R03} (\href{https://heroapps.io}{https://heroapps.io}) for creating, populating, and running models using a web browser.

\textbf{dampack} is a decision analytic modelling package providing a suite of functions for analyzing and visualizing the health economic outputs of mathematical models. This package is the closest in terms of requirements to \textbf{BCEA}.

\section{Cost-effectiveness models}
\begin{figure}
\includegraphics[width=0.9\columnwidth]{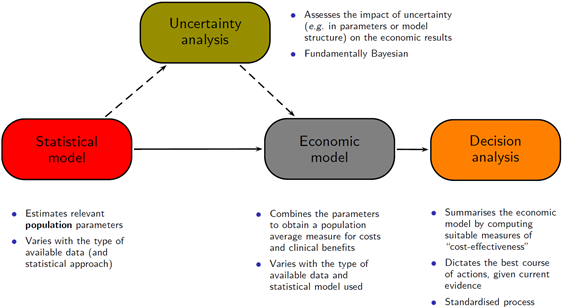}
\caption{The workflow for the health economic cost-effectiveness analysis and decision making process.} \label{fig:workflowHE}
\end{figure}

Decision making in health economics aims to identify the best course of action given the model specification and current evidence.
Using a decision-theoretic framework in the presence of uncertainty ensures rational decision-making \citep{Claxton1999}.
The process of cost-effectiveness analysis begins by describing the uncertainty on unknown quantities. Then, for each intervention, the cost and benefits are given a value by means of a utility measure. We finally select the most cost-effective intervention associated with the maximum expected utility.
From Figure~\ref{fig:workflowHE} we see how uncertainty analysis is needed in order to assess the impact of the uncertainty on the economic results \cite{Briggs2012}.
This process is fundamentally a Bayesian one.
An alternative frequentist approach, e.g using bootstrapping, may not capture the potential correlations. A joint marginal model may do better but is not realistic when the distributions of benefits may be bounded and costs skewed.

More formally, health economics is typically concerned with evaluating a set of interventions $t\in\mathcal{T}=(0,1,\ldots,T)$ that are available to treat a given condition. These may be drugs, life-style modification or complex interventions --- the general concepts of economic evaluations apply regardless. We only consider here the most important issues; comprehensive references include \citep{WillanBriggs:2006,Briggsetal:2006,Baio:2012}.

As mentioned above, the economic outcome is a multivariate response $y=(e,c)$, represented by a suitable clinical outcome (e.g.\ blood pressure or occurrence of myocardial infarction), together with a measure of the costs associated with the given intervention. On the basis of the available evidence (e.g.\ coming from a randomised controlled trial or, more likely, a combination of different sources, including observational data), the problem is to decide which option is ``optimal'' and should then be applied to the whole homogeneous population. In the context of publicly funded health-care systems (such as those in many European countries, including the UK National Health Service, NHS), this is a fundamental problem as public resources are finite and limited and thus it is often necessary to prioritise the allocation of public funds on health interventions.

Crucially, ``optimality'' can be determined by framing the problem in decision-theoretic terms \citep{O'HaganStevens:2001,Spiegelhalteretal:2004,Briggsetal:2006,Baio:2012}, which implies the following steps.
\begin{itemize}
\item Characterise the \textit{variability} in the economic outcome $(e,c)$, which is typically due to sampling, using a probability distribution $p(e,c\mid\boldsymbol\theta)$, indexed by a set of parameters $\boldsymbol\theta$. Within the Bayesian framework, \textit{uncertainty} in the parameters is also modelled using a probability distribution $p(\boldsymbol\theta)$. 
\item Value the consequences of applying a treatment $t$, through the realisation of the outcome $(e,c)$ by means of a \textit{utility function} $u(e,c;t)$. 
\item Assess ``optimality'' by computing for each intervention the expectation of the utility function, with respect to both ``population'' (parameters) and ``individual'' (sampling) uncertainty/variability
\[ \mathcal{U}^t = \E\left[u(e,c;t)\right].\]
In line with the precepts of (Bayesian) decision theory, \textbf{\textit{given current evidence}} the ``best'' intervention is the one associated with the maximum expected utility. This is because it can be easily proved that maximising the expected utility is equivalent to maximising the probability of obtaining the outcome associated with the highest (subjective) value for the decision-maker \citep{BernardoSmith:1999,Briggsetal:2006,Lindley:2006,Baio:2012}. 
\end{itemize}
Under the Bayesian framework, $\mathcal{U}^t$ is dimensionless, i.e. it is a pure number, since both sources of basic uncertainty have been marginalised out in computing the expectation. Consequently, the expected utility allows a direct comparison of the alternative options. 

While the general setting is fairly straightforward, in practice, the application of the decision-theoretic framework for health economic evaluation is characterised by the following complications.
\begin{enumerate}
\item As any Bayesian analysis, the definition of a suitable probabilistic description of the current level of knowledge in the population parameters may be difficult and potentially based on subjective judgement. 
\item There is no unique specification of the method of valuation for the consequences of the interventions is (i.e. what utility function should be chosen):
\item Typically, replacing one intervention with a new alternative is associated with some risks such as the irreversibility of investments \citep{Claxton:1999b}. Thus basing a decision on current knowledge may not be ideal, if the available evidence-base is not particularly strong/definitive. 
\end{enumerate}

As for the utility function, health economic evaluations are generally based on the
\textit{(monetary) net benefit} \citep{StinnettMullahy:1998} 
\begin{eqnarray*}
\label{eq:netbenefit}
u(e,c;t) = ke - c.
\end{eqnarray*}
Here $k$ is a willingness-to-pay parameter, used to put cost and benefits on the same scale and represents the budget that the decision-maker is willing to invest to increase the benefits by one unit. The main appeal of the net benefit is that it has a fixed form, once the variables $(e,c)$ are defined, thus providing easy guidance to valuation of the interventions. Moreover, the net benefit is linear in $(e,c)$, which facilitates interpretation and calculations. Nevertheless, the use of the net benefit presupposes that the decision-maker is \textit{risk neutral}, which is by no means always appropriate in health policy problems \citep{Koerkampetal:2007}.  

If we consider the simpler scenario where $\mathcal{T}=(0,1)$, decision-making can be equivalently effected by considering the {\em expected incremental benefit\/} (of treatment $1$ over treatment $0$)
\begin{eqnarray}
\label{eq:einb}
\mbox{EIB} = \mathcal{U}^1 - \mathcal{U}^0
\end{eqnarray}
if $\text{EIB} > 0$,
then $\mathcal{U}^1 > \mathcal{U}^0$ and therefore $t=1$ is the optimal treatment (being associated with the highest expected utility). 

In particular, using the monetary net benefit as utility function, (\ref{eq:einb}) can be re-expressed as
\begin{eqnarray}
\mbox{EIB} = \E[k\Delta_e - \Delta_c] = k\E[\Delta_e] - \E[\Delta_c] \label{EIB}
\end{eqnarray}
where 
\[
\Delta_e = \E[e \mid \boldsymbol \theta^1] - \E[e \mid \boldsymbol\theta^0] = \mu^1_e -\mu^0_e
\] 
is the average increment in the benefits (from using $t=1$ instead of $t=0$) and similarly 
\[\Delta_c=\E[c\mid \boldsymbol\theta^1] - \E[c\mid\boldsymbol\theta^0] = \mu^1_c -\mu^0_c\] 
is the average increment in costs deriving from selecting $t=1$.

If we define the \textit{Incremental Cost-Effectiveness Ratio} as
\[
\mbox{ICER}= \frac{\E[\Delta_c]}{\E[\Delta_e]}
\]
then it is straightforward to see that when the net monetary benefit is used as utility function, then 
\[
\mbox{EIB} > 0 \quad \mbox{if and only if} \quad
\left\{
\begin{array}{lr}
\displaystyle k > \frac{\E[\Delta_c]}{\E[\Delta_e]}=\mbox{ICER}, & \quad \mbox{for $\mbox{E}[\Delta_e]>0$} \\[10pt]
\displaystyle k < \frac{\E[\Delta_c]}{\E[\Delta_e]}=\mbox{ICER}, & \quad \mbox{for $\mbox{E}[\Delta_e]<0$}
\end{array} \right.
\]
and thus decision-making can be equivalently effected by comparing the ICER to the willingness-to-pay threshold. 

Notice that, in the Bayesian framework, $(\Delta_e, \Delta_c)$ are random variables, because while sampling variability is being averaged out, these are defined as functions of the parameters $\boldsymbol\theta = (\boldsymbol\theta^1, \boldsymbol\theta^0)$. The second layer of uncertainty (i.e. the population, parameters domain) can be further averaged out.  Consequently, $\E[\Delta_e]$ and $\E[\Delta_c]$ are actually pure numbers and so is the ICER.

% risk aversion
The utility for the decision maker is assumed above to be described by the monetary net benefit.
This assumption imposes a form of risk neutrality on the decision maker,
which might not be always reasonable. A scenario considering risk aversion
explicitly, with different risk aversion profiles, can be implemented by
extending the form of the utility function. One of the possible ways to
include the risk aversion in the decision problem is to re-define the utility
function \label{eq:netbenefit} as:
$$
u(b,r) = \frac{1}{r} [1 - \exp(-rb)]
$$
where the parameter $r > 0$ represents the risk aversion attributed to the
decision maker. The higher the value of $r$, the more risk-averse the decision
maker is considered to be, where $b := ke - c$ is the monetary net benefit.

% mixed strategies
We have assumed that technologies and interventions would be used in total isolation,
i.e. they would completely displace each other when chosen. In reality this
happens rarely, as new interventions are not completely implemented for all
patients in a certain indication. In general, the previous strategies usually
maintain a share of the market over time. This is due to a number of factors,
for example resistance to the novel intervention or preference of use of
different technologies in different patients.
When the market shares of the other available technologies cannot be set
to zero, the expected utility in the overall population can be computed as a
mixture:
\[
\bar{\mathcal{U}}=\sum_{t=0}^T q_t \mathcal{U}^t = q_0 \mathcal{U}^0 + q_1 \mathcal{U}^1 + \cdots + q_T \mathcal{U}^T
\]
with $q_t\geq 0 \; \forall t \in \{0,\cdots,T\}$ and $\sum_t q_t = 1$. For each intervention $t$, the quantity $q_t$ represents its market share and $\mathcal{U}^t$ its expected utilities. The resulting quantity $\bar{\mathcal{U}}$ can be easily compared with the ``optimal'' expected utility $\mathcal{U}^*$ to evaluate the potential losses induced by the different market composition. In other terms, the expected utility for the chosen market scenario is the weighted average of the expected utility of all treatment options $t$ with the respective market share $q_t$ as weights.
Although an established part of decision theory, risk aversion and mixed strategies are not yet prevalent in the health economics cost-effectiveness analysis literature.

% EVI
While uncertainty in the utility function is irrelevant for decision-making \cite{Claxton:1999b}, uncertainty plays a key role in determining whether the current evidence is sufficient to support decision-making or whether the decision should be deferred until additional evidence can be collected. This deferral decision can be based on a concept known as value of information (VoI).
In an HTA context, a VoI analysis calculates the potential value of collecting additional information to improve decision making. If information has high value, then it may be advisable to collect additional information before determining the optimal treatment.
There are two VoI measures available in the \textbf{BCEA} package: Expected Value of Perfect Information (EVPI) and Expected Value of Partial Perfect Information (EVPPI).
The EVPI quantifies the economic ``cost'' of the uncertainty in the cost-effectiveness evaluation. This quantification is based on the Opportunity Loss (OL), which measures the potential losses caused by choosing the most
cost-effective intervention on average when there is a chance it may not be the intervention
with the highest utility. This occurs when the model uncertainty results in some settings where the current optimal treatment is non-optimal. EVPI calculates the value of knowing the exact value of the utilities for the different interventions. This would allow the decision makers to know the optimal treatment with certainty.
We assume that we have "known-distribution" utility, which only includes parameter uncertainty defined as $u(e(\theta),c(\theta)$.
The EVPI is defined as follow.
$$
\mbox{EVPI} = \mathbb{E}[OL(\theta)] = \mathbb{E}_{\theta}\left[\max_d\{u(e(\theta),c(\theta),d)\}\right] - \max_d\{\mathbb{E}_{\theta}\left[u(e(\theta),c(\theta),d)\right]\}  
$$
In the case where the EVPI is high compared to the cost of any proposed additional research, it is useful to know where to target that research to reduce the decision uncertainty efficiently.
EVPPI is the value of learning the exact value of a single parameter or a specific set of parameters in the economic model.
EVPPI is defined as
$$
\mbox{EVPPI}(\varphi) = \mathbb{E}_{\varphi} \left[ \max_d \{\mathbb{E}_{\theta|\varphi}\left[u(e(\theta),c(\theta),d)\right]\}\right] - \max_d\{\mathbb{E}_{\theta}\left[u(e(\theta),c(\theta),d)\right]\} 
$$
Parameter subsets with high value should be prioritised for investigation in future research.

\section{Overview of functionality and API}
The package structure is purposefully straightforward as it is intended to be used by health economic modellers who may not necessarily be familiar with using R.
They may be more familiar with using MS Excel or TreeAge \citep{TreeAge}, for example.

\subsection{Design principles}

\begin{enumerate}
    \item Simple: There is only one call to \texttt{bcea()} required and then a call to which ever plotting type is required. The data manipulation and calculation is all handled by \texttt{bcea()} and the interface is minimal.
    \item Consistent: All plotting functions take a bcea-class object as input. The additional argument to modify the look of the plot are similar between plot types.
    \item Flexible: Plots can be intuitively modified in multiple way. Also, if the user wishes to partially modify the analysis and repeat the steps they can use helper functions to e.g. change the comparator intervention or change the willingness to pay. 
\end{enumerate}

If we were to create multiple plots it would be wasteful to recompute common, unchanged statistics.
It is also not software engineering best practice to have a single function responsible for both handling data and logic.
Computation time is not a main issue with \textbf{BCEA} since the modelling, which is usually the most computationally intensive step in a CEA, has been performed prior to using it. So computing many output statistics that may not be eventually used incurs little over-head.
Therefore, we separated the plotting from the data manipulation following the principle of decoupling.
That is, many of the data structures required for creating plots in \textbf{BCEA} have been preprocessed in \texttt{bcea()}.

In the case where the calculation of a particular output does have some non-negligible run-time there are separate functions, such as for EVPPI. This is more of an issue if we perform multiple analyses. We will return to this in the examples.

\subsection{Cost-effectiveness inputs}
\textbf{BCEA} accepts as inputs the outcomes of a CEA comparing
different interventions or strategies, ideally but not necessarily produced
using MCMC (Markov Chain Monte Carlo) methods.

In general, \textbf{BCEA} requires multiple simulations (often in the order of tens of thousands) from an economic model that
compares at least two different interventions based on their overall cost and effectiveness
measures to produce the standardised output.
The effectiveness, or efficacy, measure can be given in any form, be it a hard
outcome (e.g. number of avoided cases) or a soft outcome (e.g. QALYs, Quality-Adjusted Life Years).

Thus, the minimum input which must be given to \textbf{BCEA} is composed of
two $n_{sim} \times n_{int}$ matrices, where $n_{sim}$ is the number of simulations used to
perform the analysis (at least 2 but ideally >1000 \citep{Hatswell2018}) and $n_{int}$ is the number of interventions being compared (again, at least 2 are required) . These two matrices contain all the basic information needed by \textbf{BCEA} to perform a health economic comparison
of the alternative interventions.
We assume, in general, that the statistical model underlying the economic
analysis is performed in a fully Bayesian framework. This implies that the simulations for the economic multivariate outcome $(e, c)$ are in fact from the relevant posterior distributions. This need not be the case however and any PSA random sample could be used.

\subsection{The bcea class}
Analyses are centred around the use of the result of running \texttt{bcea()}.
This is the function which creates a list of preprocessed values to use in the various plotting and other functions.
\texttt{bcea()} is a wrapper function for a collection of functions which calculate common cost-effectiveness analysis statistics.
These are all named using the convention \texttt{compute\_*()}.
Table~\ref{tab:bcea_fns} gives the list of the internal cost-effectiveness analysis functions used in \texttt{bcea()}.
The the names of these outputs in the returned list from running \texttt{bcea()} are given in the Name column.

\begin{table}[!h]
\begin{center}
 \begin{adjustwidth}{-2cm}{}
\begin{tabular}{ r l c l } 
 \hline
Method & Description & Equation & Name\\
 \hline
 \texttt{compute\_CEAC()} & Cost-Effectiveness Acceptability Curve & $p(IB > 0 | k)$ & \texttt{ceac} \\
 \texttt{compute\_EIB()} & Expected Incremental Benefit & $\mathcal{U}^1 - \mathcal{U}^0$ & \texttt{eib} \\
 \texttt{compute\_EVI()} & Expected Value of Information & $\E[\mathcal{U}^*(\theta) - \mathcal{U}^*]$ & \texttt{evi} \\
 \texttt{compute\_IB()} & Incremental benefit & $k \Delta e - \Delta c$ & \texttt{ib} \\ 
 \texttt{compute\_ICER()} & Incremental cost-effectiveness ratio & $\Delta c/\Delta e$ & \texttt{ICER}\\ 
 \texttt{compute\_kstar()} & Optimal willingness-to-pay, $k^*$ & $\min\{k : IB < 0 \}$ & \texttt{kstar} \\ 
 \texttt{compute\_ol()} & Opportunity Loss & $\mathcal{U}^*(\theta) - \mathcal{U}(\theta^\tau)$ & \texttt{ol} \\
 \texttt{compute\_U()} & Expected utility for each simulation & $\E[\mathcal{U}; n]$ & \texttt{U} \\
 \texttt{compute\_Ustar()} & Maximum utility among the comparators & $\max \{i : \mathcal{U}^i \geq \mathcal{U}^j \}$ & \texttt{Ustar} \\
 \texttt{compute\_vi()} & Value of Information & $\mathcal{U}^*(\theta) - \mathcal{U}^*$ & \texttt{vi} \\
 \texttt{compute\_p\_best\_interv()}$^{\dagger}$ & Probability best intervention & $p(\mathcal{U}^i \geq \mathcal{U}^j)$ & \texttt{p\_best\_interv} \\
 \texttt{compute\_ceaf()}$^{\dagger}$ & Cost-effectiveness acceptability frontier & $\max \{ p(IB^i > 0); i\}$ & \texttt{ceaf} \\
 \hline
\end{tabular}
\caption{\label{tab:bcea_fns}\texttt{bcea()} component functions and details from the \textbf{BCEA} package. $^{\dagger}$ additional value appended using \texttt{multi.ce()}}
\end{adjustwidth}
\end{center}
\end{table}

Other than the objects listed in Table~\ref{tab:bcea_fns}, \texttt{bcea()} also returns values of intermediate and ancillary objects, that have been used to calculate key statistics or will be useful when plotting.

The user is free to use the results of \texttt{bcea()} independent of other \textbf{BCEA} functions, of course.
Statistics can be accessed in the usual ways for named lists, e.g using the basic extraction operator \texttt{\$}.

\subsection{Plotting functions}
There are a numerous plotting functions in \textbf{BCEA}.
Each plot type can be rendered using base R plotting, \textbf{ggplot2} and \textbf{plotly}.
Each main plotting function uses an approach taken from the {\it strategy pattern} (also known as the policy pattern) \citep{GoF}.
Instead of implementing a single algorithm directly, code receives run-time instructions as to which in a family of algorithms to use.
In our case, this is which graphical device to use.
Therefore, 'under the hood' each top-level plotting function has associated *\texttt{\_base()}, *\texttt{\_ggplot()} and *\texttt{\_plotly()}.
This is not strictly a strategy pattern because R it is an interpreted language and the S3 object system in which \textbf{BCEA} is written is a generic-function style of object oriented programming.
The top level functions serve to validate and prepare the input data so in principle the lower-level functions could be accessed directly but this is not recommended.
There is no noticeable run-time impact with this approach.

Common functionality between types of plots is extracted out for improved maintenance and reliability.
For instance, positioning the legend with \texttt{where\_legend()} or setting the theme with \texttt{theme\_default()}.
This does however mean that the defaults are not tailored to each plot and the user may want to change them.
For example, the CEAC plot may have curves drawn from bottom left to top right and so wish to place the legend in the bottom right corner.
This can be achieved in the usual way with base R and \textbf{ggplot2} commands.

Table~\ref{tab:plot_fns} gives the list of the main plotting functions.

\begin{table}[!h]
\begin{center}
\begin{tabular}{ r l c } 
 \hline
Method & Description &\\
 \hline
 \texttt{plot()} & Grid of multiple plots, depending on class dispatch &  \\ 
 \texttt{ceac.plot()} & Cost-Effectiveness Acceptability Curve &  \\ 
 \texttt{ceaf.plot()} & Cost-Effectiveness Frontier &  \\ 
 \texttt{ceef.plot()} & Cost-Effectiveness Efficiency Frontier &  \\ 
 \texttt{ceplane.plot()} & Cost-Effectiveness Plane scatter plot &  \\ 
 \texttt{contour()} & Contour Plots for the Cost-Effectiveness Plane &  \\ 
 \texttt{contour2()} & Specialised Cost-Effectiveness Plane Contour Plot &  \\ 
 \texttt{eib.plot()} & Expected Incremental Benefit &  \\ 
 \texttt{evi.plot()} & Expected Value of Information  &  \\ 
 \texttt{ib.plot()} & Incremental Benefit Distribution &  \\ 
 \texttt{info.rank()} & Ranking of information value for each parameter &  \\ 
 \hline
\end{tabular}
\caption{\label{tab:plot_fns}Plotting functions in \textbf{BCEA}.}
\end{center}
\end{table}

\subsection{Other output functions}
The main strength of \textbf{BCEA} is its range of easy-to-use, flexible plotting functions but there are also several other additional functions which can be used to further explore the CEA model output and present the results.
Table~\ref{tab:misc_fns} gives the list of these functions.

Two ways of summarising the \texttt{bcea()} output are provided in \textbf{BCEA}.
The \texttt{summary()} function returns a table reporting the basic results of the health economic analysis.
In addition to these basic health economic measures, \textbf{BCEA} provides some summary measures for the PSA, allowing a more in-depth analysis of the variation observed in the results, speciffcally the CEAC and the EVPI.
\texttt{sim\_table()} produces the health economics outputs in correspondence of each simulation, including the utility values for the interventions, the maximum utility value among the comparators, the incremental benefit for the comparison between interventions, the opportunity loss and the value of information.

Standardising the presentation of the final reporting of a CEA has multiple benefits, including ease of comprehension for the reader, a shorter time to produce, consistency and fewer errors. The \texttt{make.report()} function constructs an automated report from the output of the CEA.

The expected value of information function \texttt{evppi()} returns several related outputs in a list, including a vector of EVPPI values for each value of willingness to pay, a vector of the values for the EVPI for each value of  willingness to pay, the parameters for which the EVPPI has been calculated, and the calculation method used for the EVPPI.

\begin{table}[!h]
\begin{center}
\begin{tabular}{ r l c } 
 \hline
Method & Description & \\
 \hline
  \texttt{evppi()} & Expected Value of Perfect Partial Information &  \\ 
 \texttt{make.report()} & Automated report from the cost-effectiveness analysis using RMarkdown &  \\
 \texttt{sim\_table()} & Summary table of the simulations from the cost-effectiveness analysis &  \\ 
 \texttt{summary()} & Table of summary statistics of the health economic evaluation &  \\ 
 \hline
\end{tabular}
\caption{\label{tab:misc_fns}Other output functions in \textbf{BCEA}.}
\end{center}
\end{table}

The following sections build on the basic \texttt{bcea()} implementation and allow modification of an analysis as part of the workflow.

\subsubsection{Replacing parameter values in a bcea object}
To reemphasise, the use of \textbf{BCEA} centres around the \texttt{bcea()} function.
That is, the first step in an analysis using \textbf{BCEA} is to pass the random sample of costs and health consequences from a CEA model into \texttt{bcea()} which then calculates the statistics used by other functions in the package.
So far we have implicitly assumed that this call to \texttt{bcea()} only happens once.

However, we may want to experiment to see how changing some of the arguments to \texttt{bcea()} change the different outputs.
To do this we do not need to rerun \texttt{bcea()} from scratch but simply modify the argument of interest.
This is done via the {\it replacement} function \texttt{`<-`} ability of R.
The current arguments implemented are to modify the comparison set, using \texttt{'SetComparison<-()'}, the reference group, using \texttt{'SetReferenceGroup<-()'}, and the maximum willingness-to-pay value, $K$, using \texttt{'SetKmax<-()'}.

\subsubsection{Extending a bcea object}
Similarly, we may wish to investigate the effect of different risk aversion functions on the \texttt{bcea()} outputs.
This is slightly different in that the form of the \texttt{bcea()} output is extended to include the additional cost-effectiveness analysis statistics using risk aversion values (relative to the risk-neutral case that is the default). In particular, this provides updated values for \texttt{U, Ustar, ib, eib, vi} and \texttt{evi}.
This has been implemented in the spirit of a {\it decorator} \citep{GoF}, again using replacement functions, where the original \texttt{bcea()} object is extended at run-time.
As well as recalculating the results of \texttt{bcea()} with the provided risk aversion parameter values, this appends a subclass \texttt{CEriskav} to the results of \texttt{bcea()}.
This then means that the same plotting functions can be called and R knows to dispatch first on the class \texttt{CEriskav} before \texttt{bcea}, if available.

In the same way as has been implemented with \texttt{'CEriskav<-()'}, mixed strategies can be investigated using the subclass \texttt{mixedAn} as part of a decorator-type function \texttt{'mixedAn<-()'}.
A vector of market shares associated with the interventions is assigned to \texttt{'mixedAn()'}.
The default assumes uniform distribution for each intervention. The updated cost-effectiveness statistics are for \texttt{U, ol} and \texttt{evi}.

\subsubsection{Multiple simultaneous comparisons}
So far we have only looked at paired comparisons of interventions.
That is, comparing an intervention with one other, usually a baseline or status quo.
The alternative is to consider all interventions simultaneously.
This is achieved in \texttt{bcea} using the \texttt{multi.ce()} function.
Similar to \texttt{'CEriskav<-()'} and \texttt{'mixedAn<-()'}, this extends the \texttt{bcea} object to include a probability of each intervention being optimal across all interventions at different willingness-to-pay thresholds.
However, in contrast to the other \texttt{bcea} extension functions, \texttt{multi.ce()} does not have a replacement format function since it does not require any new values in order to be called. The output of \texttt{ceac.plot()} is different depending on pairwise or simultaneous intervention comparisons, as we shall see in the examples.

\section{Examples}
We now show how to use \textbf{BCEA} with two case studies: smoking cessation and influenza vaccination. The data for these studies are provided as part of \textbf{BCEA}.

\subsection{Smoking cessation data}
We will first show how to use the smoking cessation data set to create individual plots and how to modify an existing analysis.
The data set consists of total cost and QALY (check) pairs for three interventions aimed at helping a smoker to quit: Self-help, Individual counselling, Group counselling. A fourth group is also included of no intervention.
The first step is to load the data from the \textbf{BCEA} package and create the \texttt{bcea} object.

\begin{example}
> data(Smoking, package = "BCEA")
> treats <- c("No intervention", "Self-help", "Individual counselling", "Group counselling")
> bcea_smoke <- bcea(e, c, ref = 4, interventions = treats, Kmax = 500)
\end{example}
For this example, we set the reference group (\texttt{ref}), e.g. status quo, to 4 (Group counselling) and the maximum willingness-to-pay (\texttt{Kmax}) as 500.
We will call the individual plots separately, starting with the cost-effectiveness plane.

\begin{example}
> ceplane.plot(bcea_smoke, comparison = 2, wtp = 250)
\end{example}
\begin{figure}[!h]
    \centering
    \includegraphics[width=0.7\columnwidth]{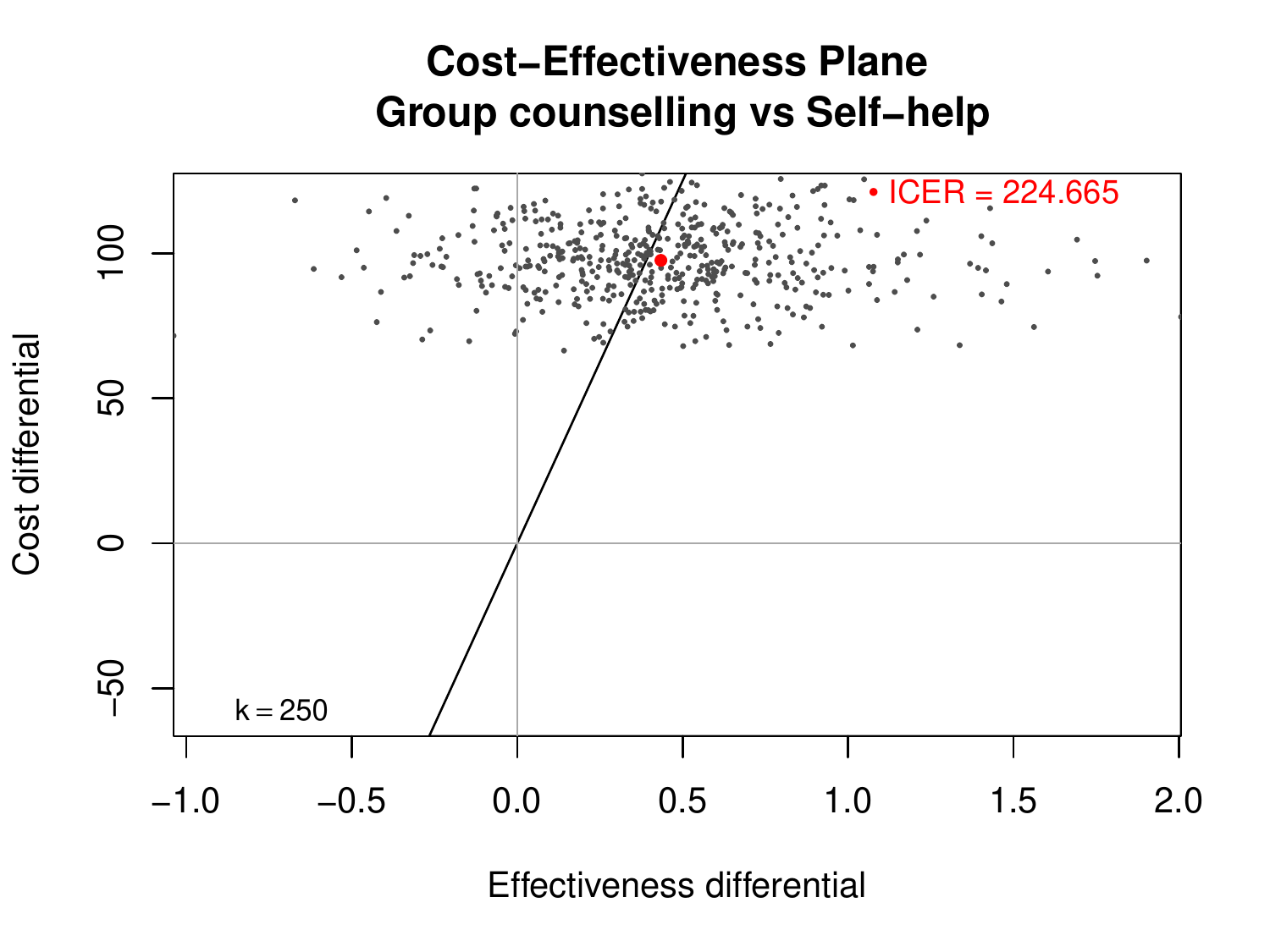}
    \caption{The cost-effectiveness plane for the Smoking example. The red dot indicates the average of the distribution of the outcomes, i.e. the ICER. The area below the diagonal line is a representation of the sustainability area, in correspondence of the fixed willingness to pay threshold, in this case fixed at 250 monetary units (the default).}
    \label{fig:mesh1}
\end{figure}

The comparison interventions are first set when \texttt{bcea()} is called and defaults to all non-reference groups but can be over-ridden when the plot is made, as in this case setting \texttt{comparison} = 2 (Self-help).
This results is a single comparison for Group counselling vs Self-help.
The plot shows a single cloud of points and so a single ICER value which is shown in the top right hand corner as $224.665$.
Also, we have set the willingness-to-pay threshold (\texttt{wtp}) at $k = 250$ indicated by the gradient of the diagonal line and the annotation at the bottom left hand corner.

We can easily modify the comparison group and then replot.
Let us change the comparison groups to include 1 (No intervention) and 3 (Individual counselling) using the \texttt{setComparison()<-} setter function.
We also create the cost-effectiveness plane plot using \textbf{ggplot2} rather than base R with the \texttt{graph} argument.
\begin{example}
> setComparisons(bcea_smoke) <- c(1,3)
> ceplane.plot(bcea_smoke, wtp = 250, graph = "ggplot2")
\end{example}
\begin{figure}[!h]
    \centering
    \includegraphics[width=0.7\columnwidth]{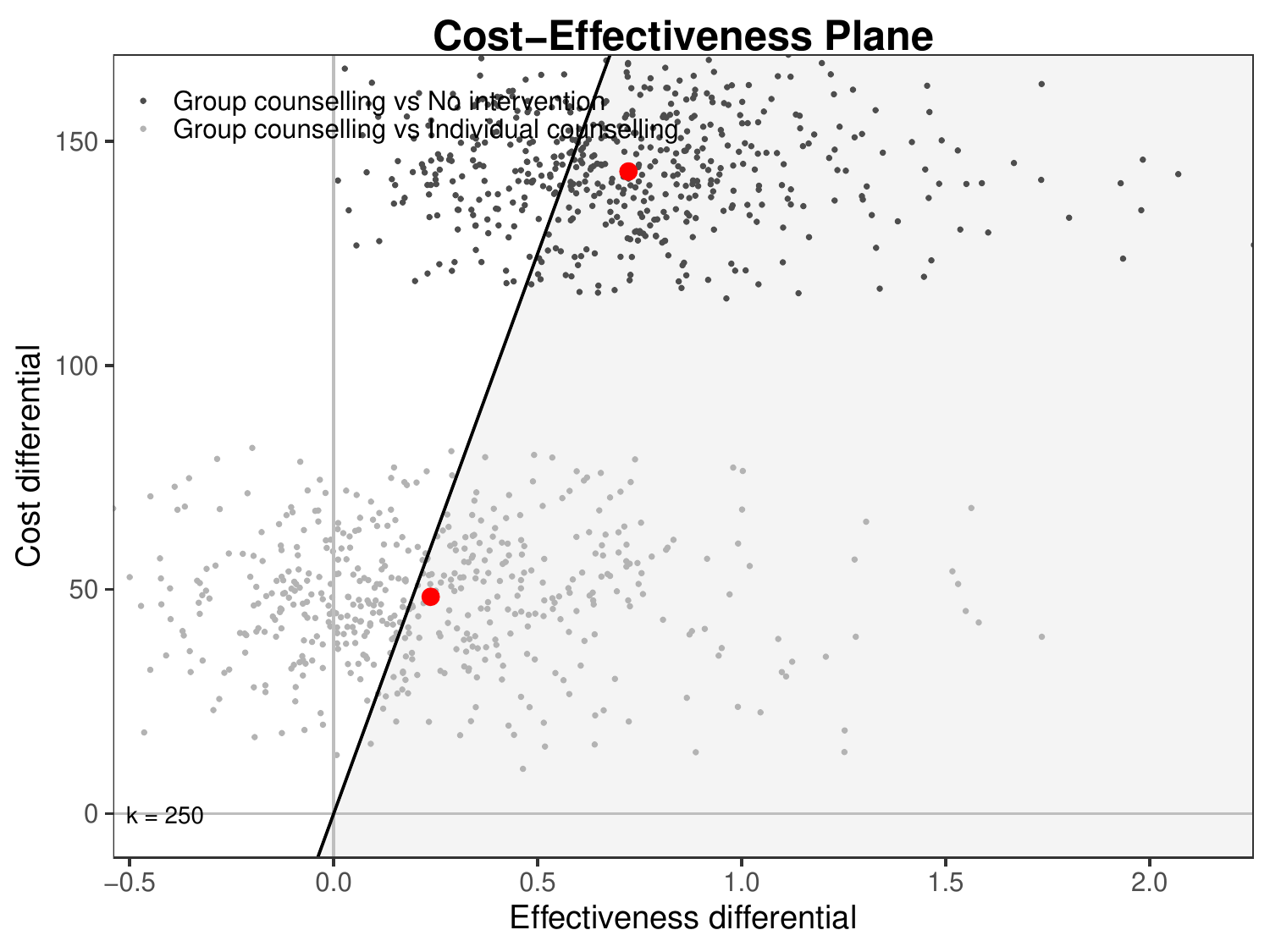}
    \caption{The cost-effectiveness plane for the Smoking example. The grey-shaded surface is a representation of the sustainability area, in correspondence of the fixed willingness to pay threshold, in this case fixed at 250 monetary units (the default).}
    \label{fig:mesh1}
\end{figure}

Now there are two clouds of points rather than one, the ICER is not shown and the information about the comparisons is moved from the title to the legend in the top left.
The same principle of plotting and modifying applies to the other plots in \textbf{BCEA}.

Next, we wish to change the analysis to make multiple simultaneous comparisons between interventions rather than paired comparisons as we saw in the previous analysis.
This is achieved as follows using the \texttt{multi.ce()} function.

\begin{example}
> bcea_smoke <- multi.ce(bcea_smoke)
\end{example}
The cost-effectiveness acceptability curve function is used in exactly the same way as for the paired comparison case.
\begin{example}
> ceac.plot(bcea_smoke)
\end{example}
\begin{figure}[!h]
    \centering
    \includegraphics[width=0.7\columnwidth]{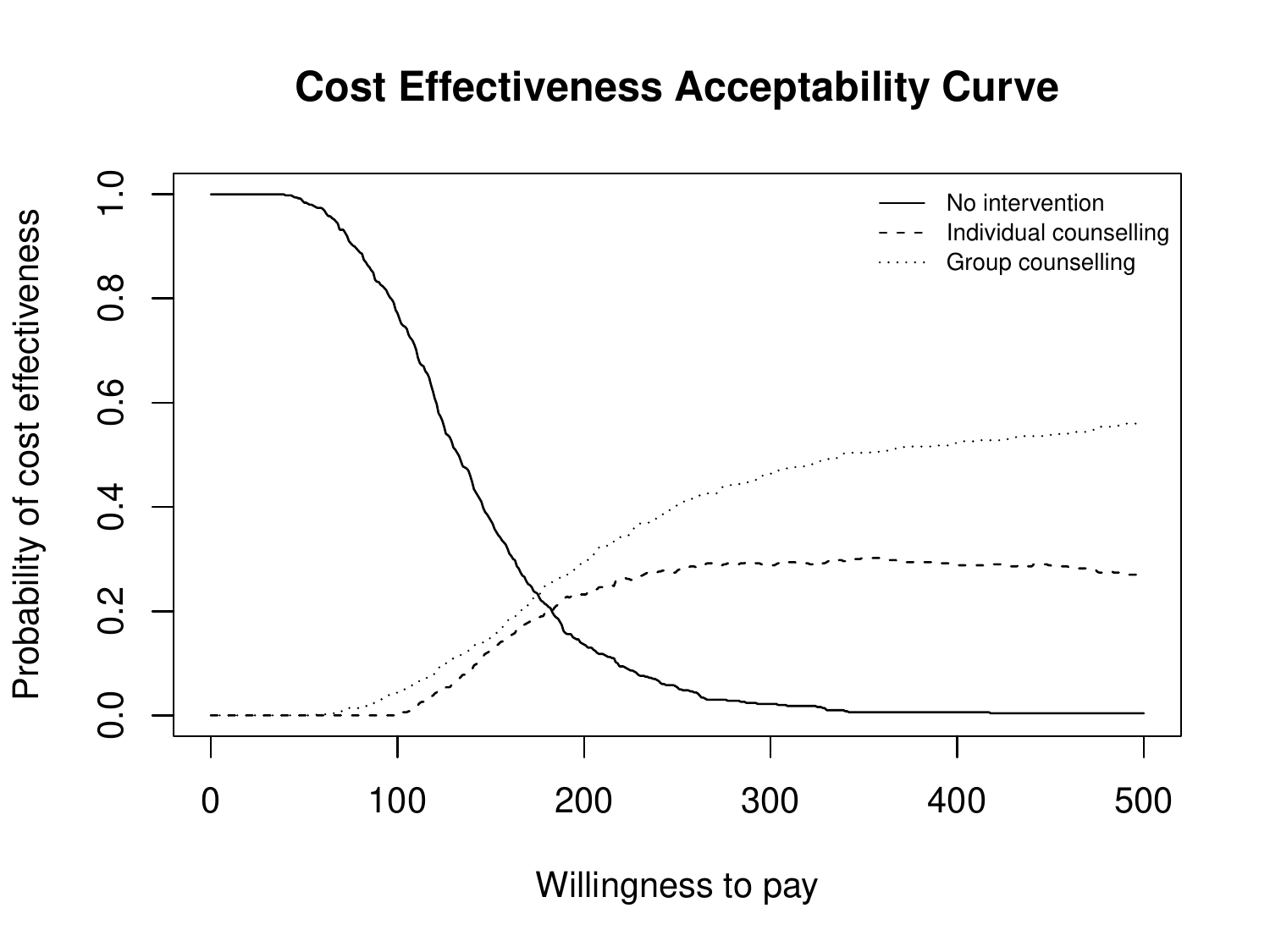}
    \caption{A graphical representation of the probability of cost-effectiveness of
each treatment when other comparators are considered. In this case three interventions are compared.
The information given is substantially
different from the pairwise CEACs, since it allows for the evaluation of the best treatment option over the considered grid of willingness to pay values. The uncertainty associated with the decision can be inferred by the distance between the treatment-specific curves.}
    \label{fig:mesh1}
\end{figure}

We see that we have three curves (rather than two which we would have in the paired comparison case) whose total area under the curves sums to one since all interventions are considered simultaneously.
Note that in this example we have chosen No intervention and Individual counselling as comparator groups and Group counselling as the reference group so these three options are plotted, and Self-help is omitted.
If we wanted to highlight this then the cost-effectiveness acceptability frontier can be explicitly plotted using \texttt{ceaf.plot(bcea\_smoke)}.

To extend the analysis to have a mixed strategy with market shares for all four interventions of $0.4, 0.3, 0.2, 0.1$ use the \texttt{mixedAn()<-} setter function. Now when we call the expected value of perfect information plot we get the appropriate plot.

\begin{example}  
> mixedAn(bcea_smoke) <- c(0.4, 0.3, 0.2, 0.1)
> evi.plot(bcea_smoke, graph = "ggplot", pos = "b")
\end{example}
\begin{figure}[!h]
    \centering
    \includegraphics[width=0.7\columnwidth]{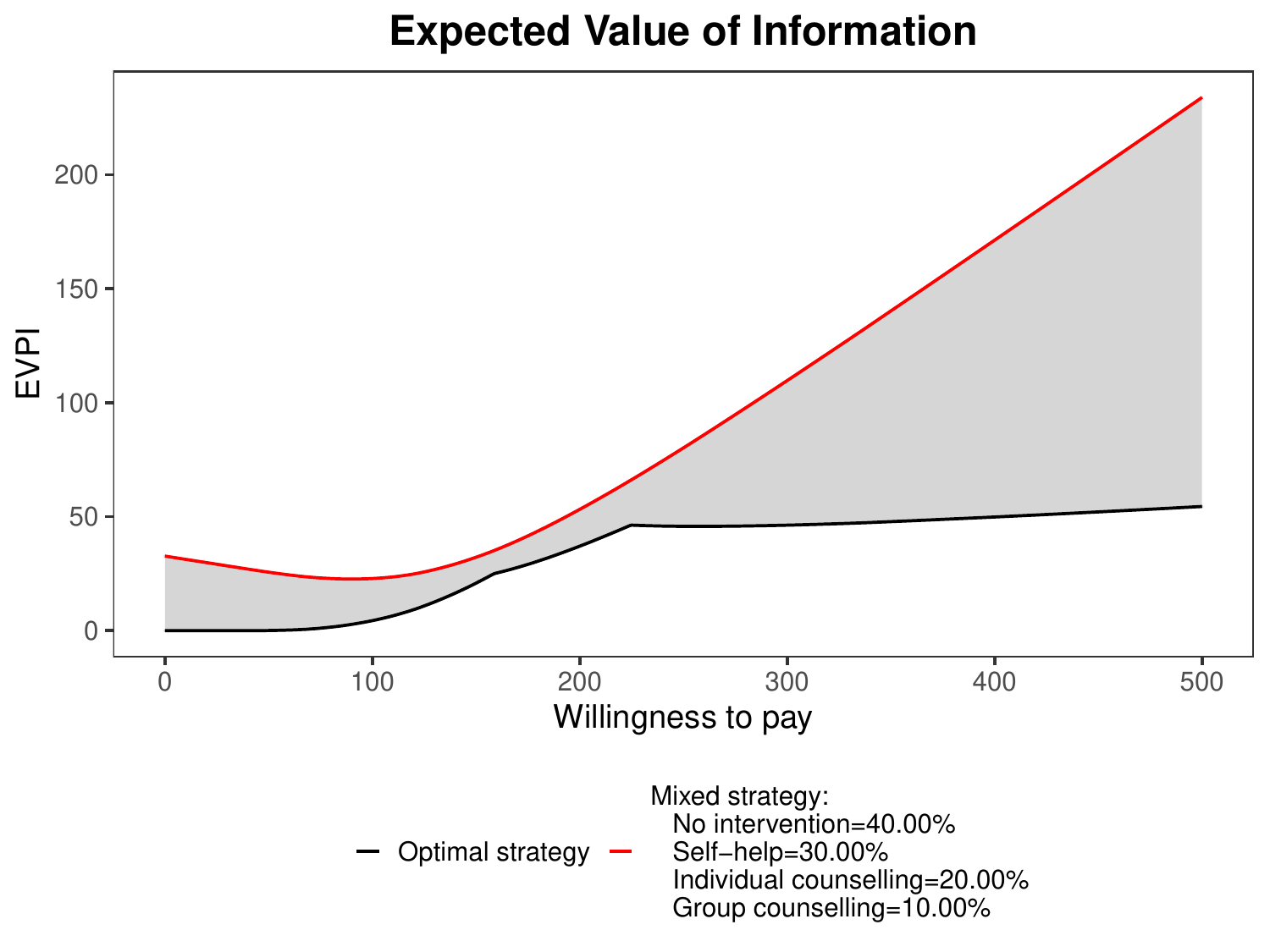}
    \caption{Represents the values of the expected value of perfect information under the optimal strategy and mixed strategy scenarios for varying willingness to pay thresholds. It is clearly shown that in this case the EVPI for the mixed strategy is always greater than for the optimal strategy, due to the sub-optimality of the market shares leading to higher values of the opportunity loss.}
    \label{fig:mesh1}
\end{figure}

Lastly, we can generalise the utility function to incorporate risk aversion for values $0, 0.005, 0.020, 0.035$.

\begin{example}
> r <- c(0, 0.005, 0.020, 0.035) 
> CEriskav(bcea_smoke) <- r
\end{example}
The wrapper function \texttt{plot()} works different in this case.
Rather than returning the 2-by-2 grid of plots for the simple \texttt{bcea} object in the non-risk aversion case, in the \texttt{mixedAn} case a call to \texttt{plot()} returns a pair of plots, for EIB and EVI incorporating the risk aversion values.

\begin{example}
> plot(bcea_smoke)
\end{example}

\begin{figure}[!h]
    \centering
    \includegraphics[width=0.6\columnwidth]{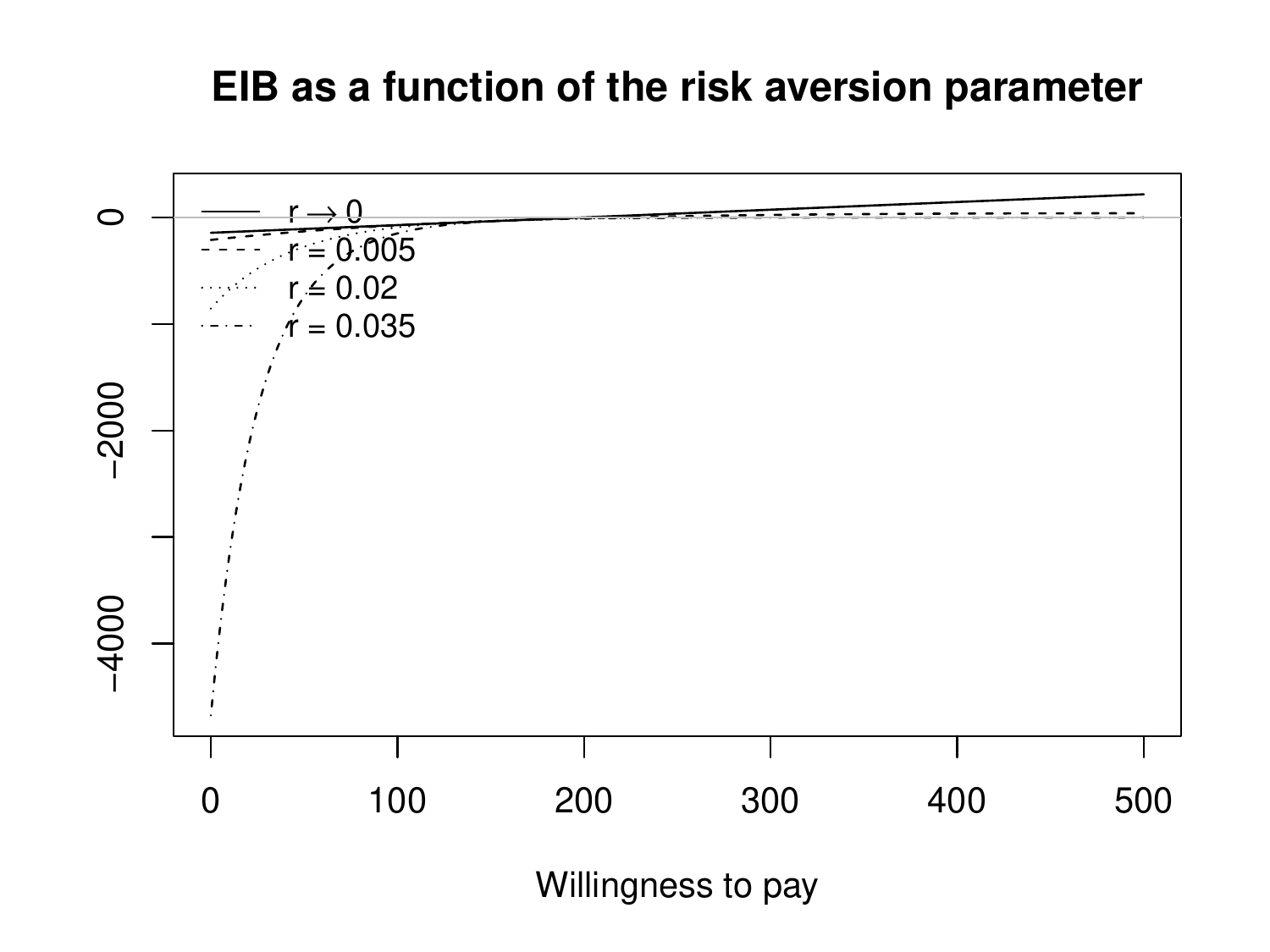}
    \includegraphics[width=0.6\columnwidth]{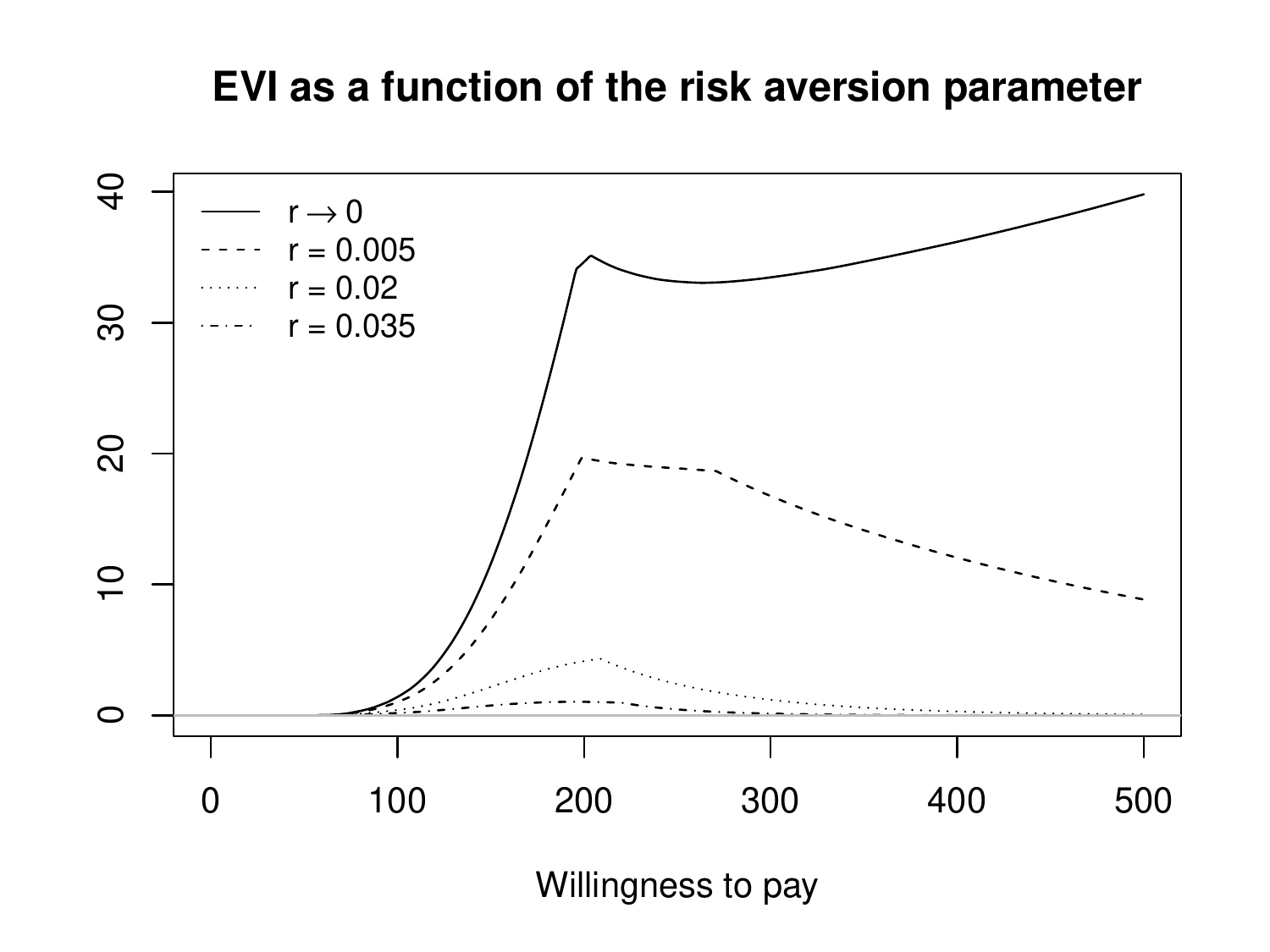}
    \caption{The Figures show the output of the plot function for the risk aversion analysis. The Figures show the effect of different risk aversion scenarios on the expected incremental benefit (EIB) and the expected value of perfect information (EVPI), respectively at the top and bottom of the Figure. It can be easily noticed that the EIB departs from linearity and the decision uncertainty represented in the EVPI grows with increasing aversion to risk.}
    \label{fig:mesh1}
\end{figure}

\subsection{Influenza vaccination data}
Next, we will use the simulated values from the vaccination study to demonstrate the expected value of information features in \textbf{BCEA}.
The vaccine data set consists of samples of total cost and QALY pairs (\texttt{c,\!\!\! e}), similar to the smoking cessation data set. However, the vaccine data set also comprises of posterior samples of model parameter values contained in the output from a model written in the JAGS software and run from R using the package \textbf{rjags}. These are what we will use to explore the value of partial information.
The vaccine analysis investigates two interventions, status quo and vaccination.
We begin, as before, by loading the data in to the current R session and running \texttt{bcea()}.

\begin{example}
> data(Vaccine, package = "BCEA")
> treats <- c("Status quo", "Vaccination")
> bcea_vacc <- bcea(e, c, ref = 2, interventions = treats)
\end{example}

The simplest way to view several different types of outputs is to use the \texttt{plot()} function to generate a grid of plots.

\begin{example}
> plot(bcea_vacc)
\end{example}
\begin{figure}[!h]
    \centering
    \includegraphics[width=0.7\columnwidth]{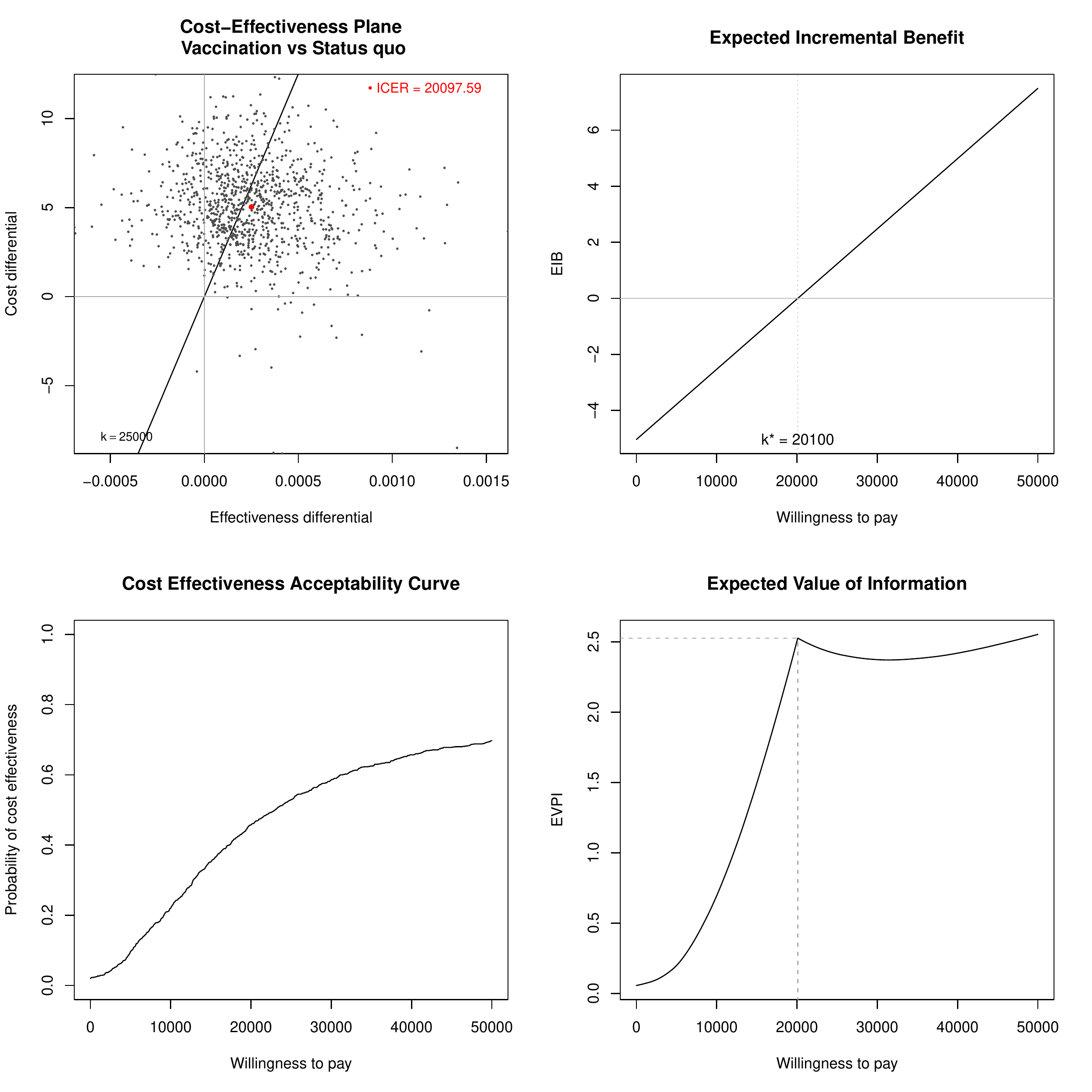}
    \caption{The summary of the health economic analysis produced by the \textbf{ggplot2} version of \texttt{plot.bcea}. The different colours and line types indicate the three pairwise comparisons versus the status quo (No intervention). The two willingness to pay values in correspondence of which the decision changes are represented in the expected incremental benefit (EIB) and expected value of perfect information (EVPI) plots. An arbitrary willingness to pay, equal to £250 per life year saved, has been chosen for the cost-effectiveness plane graph}
    \label{fig:mesh1}
\end{figure}

Specific values can be viewed using summary output functions.

\begin{example}
> summary(bcea_vacc, wtp = 10000)
\end{example}

\begin{example}
Cost-effectiveness analysis summary 

Reference intervention:  Vaccination
Comparator intervention: Status quo

Optimal decision: choose Status quo for k < 20100 and Vaccination for k >= 20100

Analysis for willingness to pay parameter k = 10000

            Expected utility
Status quo           -36.054
Vaccination          -34.826

                             EIB  CEAC  ICER
Vaccination vs Status quo 1.2284 0.529 20098

Optimal intervention (max expected utility) for k = 10000: Status quo
           
EVPI 3.0287
\end{example}

We can also pull out the sample of key cost-effectiveness analysis statistics.
Below we print the top few lines of this table.

\begin{example}
> head(sim_table(bcea_vacc, wtp = 25000)$Table)

         U1        U2        U*      IB2_1       OL         VI
1 -36.57582 -38.71760 -36.57582 -2.1417866 2.141787  -1.135907
2 -27.92514 -27.67448 -27.67448  0.2506573 0.000000   7.765431
3 -28.03024 -33.37394 -28.03024 -5.3436963 5.343696   7.409665
4 -53.28408 -47.13734 -47.13734  6.1467384 0.000000 -11.697432
5 -43.58389 -40.40469 -40.40469  3.1791976 0.000000  -4.964782
6 -42.37456 -33.08547 -33.08547  9.2890987 0.000000   2.354444
\end{example}
Let's use the value of information functions in \textbf{BCEA}.
First, below we show the EVI plot for the vaccine data.

\begin{example}
> evi.plot(bcea_vacc)
\end{example}
\begin{figure}[!h]
    \centering
    \includegraphics[width=0.7\columnwidth]{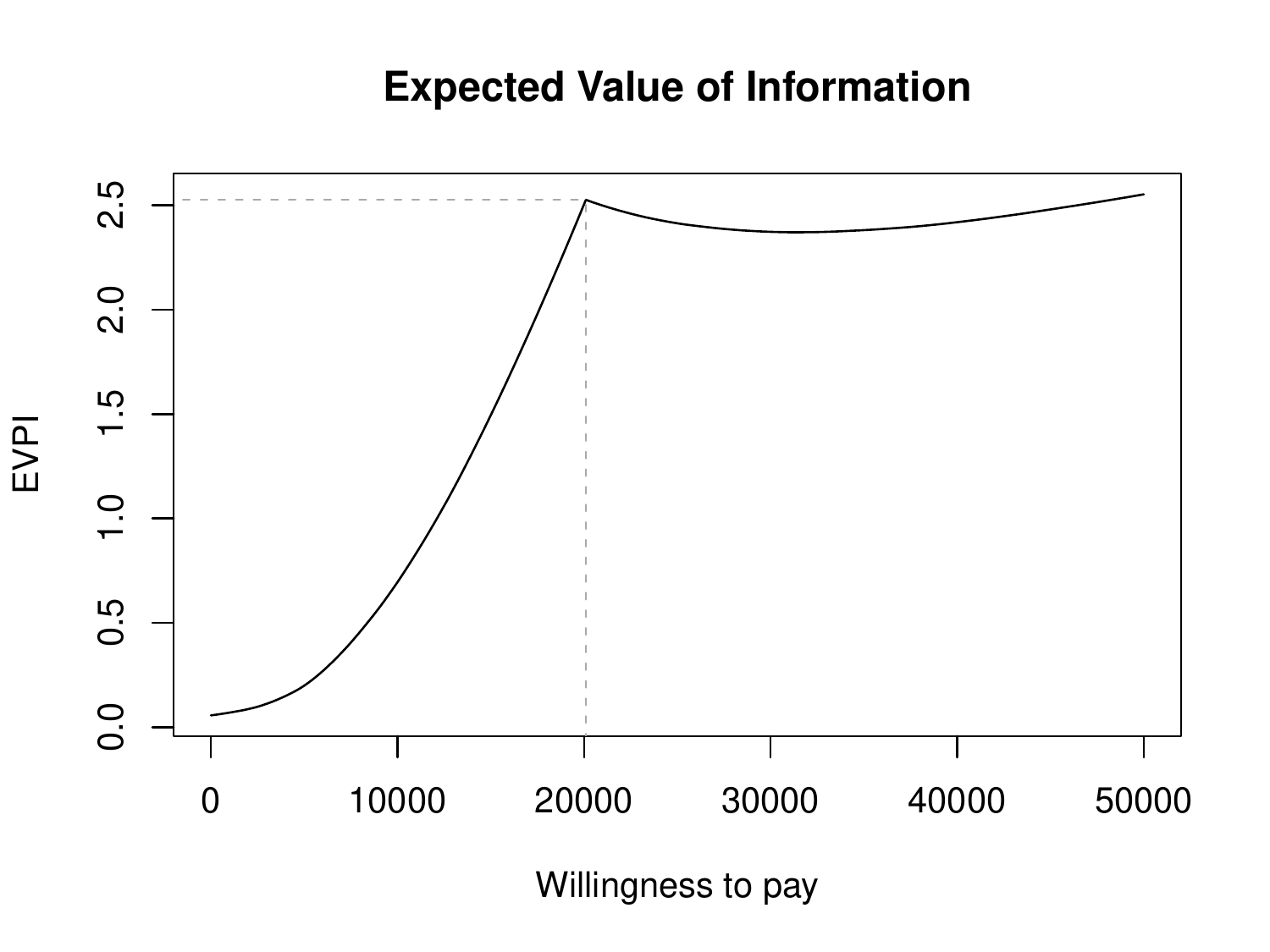}
    \caption{Expected Value of Perfect Information using vaccination data set.}
    \label{fig:mesh1}
\end{figure}

\noindent
Next, let us focus on the the specific parameters $\beta_1$ (\texttt{beta.1.}) and $\beta_2$ (\texttt{beta.2.}) to perform an EVPPI analysis.
First create the required inputs using \texttt{createInputs()} and providing the posterior samples from JAGS within the variable \texttt{vaccine\_jags}.

\begin{example}
> inp <- createInputs(vaccine_mat, print_is_linear_comb = FALSE)
> EVPPI <- evppi(bcea_vacc, c("beta.1.", "beta.2."), inp$mat)
\end{example}

\noindent
Finally, we can generate the Info-rank plot for all parameters.

\begin{example}
> info.rank(bcea_vacc, inp)
\end{example}
\begin{figure}[!h]
    \centering
    \includegraphics[width=1\columnwidth]{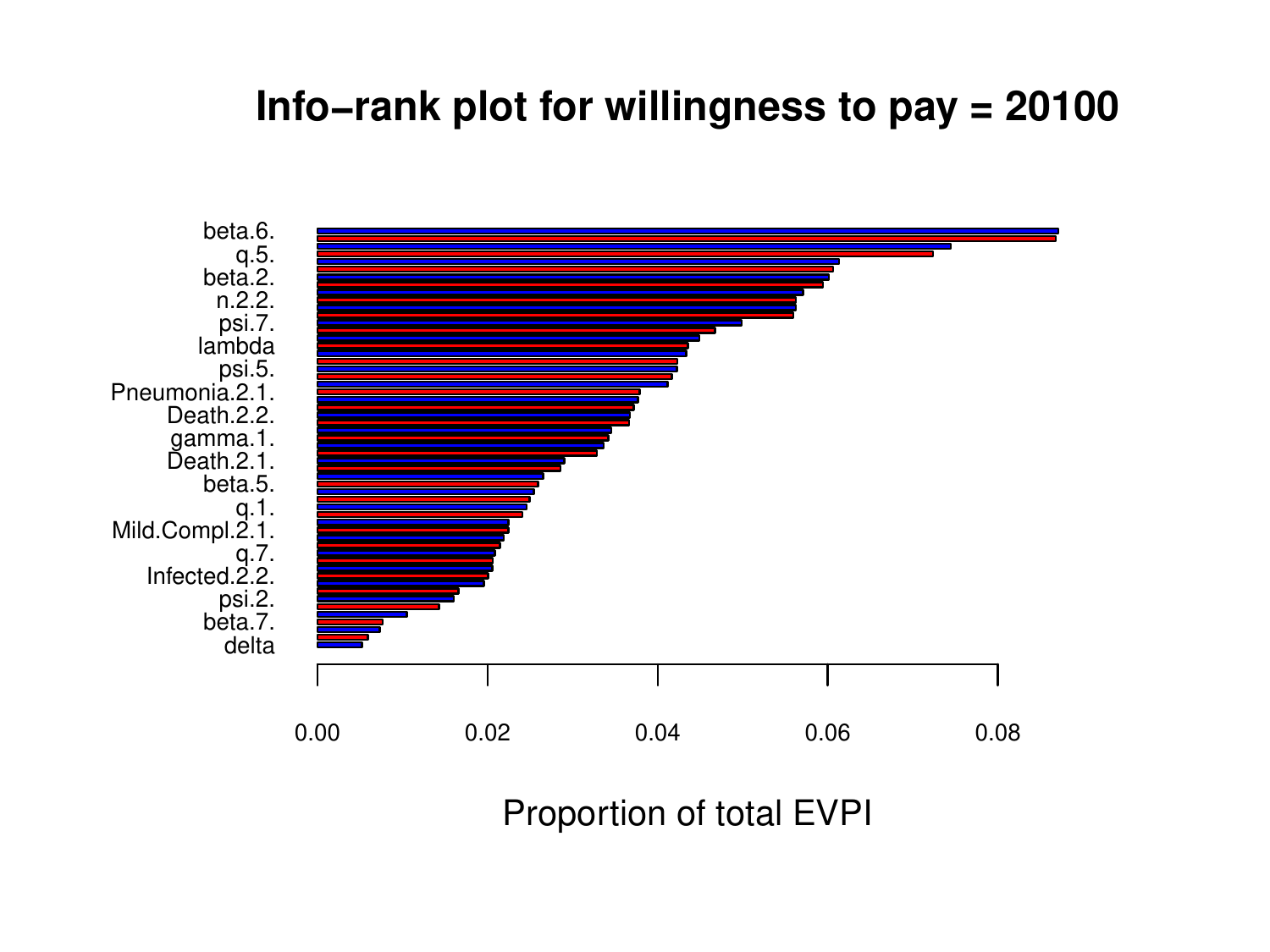}
    \caption{Info-rank plot for all model parameters in the vaccine data set and willingness to pay 20,100. Each bar quantifies the proportion of the total EVPI associated with each of the parameters used as input.}
    \label{fig:mesh1}
\end{figure}

\section{Summary}
As more and more practitioners working in health economics move away from using limited software, e.g. MS Excel, to build their models and start using R there is an increasing demand for simple to use, flexible and trustworthy packages in R. This paper introduced the \textbf{BCEA} package for this end. It has been carefully designed to allow users to perform CEA easily and consistently.
Further, a Shiny application version of \textbf{BCEA} is available called \textbf{BCEAweb} which enables users to use \textbf{BCEA} in the browser via an easy to use point-and-click interface. The web interface is available at https://egon.stats.ucl.ac.uk/projects/BCEAweb/.

\textbf{BCEA}  will continue to be refined and extended where appropriate.
\textbf{BCEA} lives inside of a fast growing ecosystem of packages designed to perform various steps in the wider CEA workflow and so future work should involve facilitating these tools easily working together and providing the full range of capabilities required.

\section{Acknowledgements}
Dr Heath is funded by Canada Research Chair in Statistical Trial Design; Natural Sciences and Engineering Research Council of Canada (award No. RGPIN-2021-03366).

\bibliography{BCEA_RJwrapper}

\address{Nathan Green\\
  Department of Statistical Science, UCL\\
  Torrington Place, UK\\
  (ORCiD 0000-0003-2745-1736)\\
  \email{n.green@ucl.ac.uk}}

\address{Anna Heath\\
  Child Health Evaluative Sciences, The Hospital for Sick Children, Toronto, ON, Canada\\
  Division of Biostatistics, Dalla Lana School of Public Health, University of Toronto, Toronto, ON, Canada\\
  Department of Statistical Science, University College London, London, UK.\\
  (ORCiD 0000-0002-7263-4251)\\
  \email{anna.heath@sickkids.ca}}
 
\address{Gianluca Baio\\
  Department of Statistical Science, UCL\\
  Torrington Place, UK\\
  (ORCiD 0000-0003-4314-2570)\\
  \email{g.baio@ucl.ac.uk}}
  
\end{article}

\end{document}